\documentclass[12pt,twoside]{article} 
\usepackage[super,sort,comma]{natbib}

\usepackage{fancyhdr}		

\newcommand{\captionv}[3]{\begin{center}\parbox{#1cm}{\caption[#2]{{\sf #3}}}
        \end{center}}

\usepackage[section]{placeins}   %

\usepackage{graphicx}

\makeatletter \renewcommand\@biblabel[1]{$^{#1}$} \makeatother
\newcommand{\cen}[1]{\begin{center} #1 \end{center}}

\usepackage[mathlines]{lineno}

\usepackage{hyperref}
\hypersetup{ colorlinks,
    citecolor=blue,
    filecolor=blue,
    linkcolor=blue,
    urlcolor=blue
}

\usepackage{xcolor}

\definecolor{gray}{rgb}{0.6,0.6,0.6}
\definecolor{red}{rgb}{0.85,0,0}
\definecolor{green}{rgb}{0,0.85,0}
\definecolor{blue}{rgb}{0,0,0.85}
\definecolor{beige}{rgb}{0.92,0.87,0.78}
\usepackage[all]{hypcap}    

\usepackage{amsmath,amssymb,amsfonts}
\usepackage{algorithmic}
\usepackage{graphicx}
\usepackage{textcomp}
\usepackage{multicol, multirow}
\usepackage{colortbl}
\usepackage{float}

\definecolor{yellow}{rgb}{0.95, 0.91, 0.48}
\definecolor{orange}{rgb}{0.97,0.77,0.50}
\definecolor{blue}{rgb}{0.60,0.80,0.97}
\definecolor{green}{rgb}{0.39,0.83,0.07}
\DeclareMathOperator*{\argminB}{argmin} 

\begin{document}

\cen{\sf {\Large {\bfseries A Model-Based Dictionary Approach for \\ Magnetic Nanoparticle Signal Prediction} \\ 
\vspace*{10mm}
Asli Alpman\textsuperscript{1,2,3}, Mustafa Utkur\textsuperscript{1,2,4}, and Emine Ulku Saritas\textsuperscript{1,2}} \\
\vspace*{10mm}
\textsuperscript{1}Department of Electrical and Electronics Engineering, Bilkent University, Ankara, 06800, Turkey\\
\textsuperscript{2}National Magnetic Resonance Research Center (UMRAM), Bilkent University, Ankara, 06800, Turkey\\
\textsuperscript{3}Department of Electrical Engineering and Computer Sciences, University of California, Berkeley, CA, 94720, USA\\
\textsuperscript{4}Department of Radiology, Boston Children’s Hospital, Harvard Medical School, Boston, MA, 02115, USA
\vspace{5mm}\\
}

\pagenumbering{roman}
\setcounter{page}{1}
\pagestyle{plain}
Author to whom correspondence should be addressed.  email: alpman@ee.bilkent.edu.tr

\begin{abstract}
\noindent {\bf Background:} 
Magnetic particle imaging (MPI) is a tracer-based medical imaging modality that enables quantification and spatial mapping of magnetic nanoparticle (MNP) distribution. The magnetization response of MNPs depends on both experimental conditions such as drive field (DF) settings and viscosity of the medium, and the magnetic parameters of MNPs such as magnetic core diameter, hydrodynamic diameter, and magnetic anisotropy constant.\\
\\
\noindent {\bf Purpose:} 
A comprehensive understanding of the magnetization response of MNPs can facilitate the optimization of DF and MNP type for a given MPI application, without the need for extensive experimentation. In this work, we propose a calibration-free
iterative algorithm using model-based dictionaries for MNP signal prediction at untested settings. \\
\\
\noindent {\bf Methods:}
Dictionaries were constructed with the MNP signals simulated using the coupled Brown–Néel rotation model.
Based on the available measurements, the proposed algorithm jointly estimates the dictionary weights and the transfer functions due to non-model based dynamics. These dynamics include the system response of the measurement setup as well as the magnetization dynamics not accounted for by the employed coupled Brown-Néel rotation model. 
First, the performance of the proposed algorithm was validated using synthetic signals generated for a known set of MNP parameters, at two different signal-to-noise ratio (SNR) levels of 1 and 10.
Next, experiments were performed on an in-house arbitrary waveform magnetic particle spectrometer (MPS) setup at 6 different viscosity levels within the biologically relevant range of 0.89 to 15.33~mPa$\cdot$s and DF frequencies between 0.25 to 2~kHz, for two different commercially available MNPs. To assess the signal prediction performance, MPS measurements from one of the viscosity levels were left out entirely and then predicted with the weight vector and transfer functions estimated with the remaining measurements. Quantitative evaluation metrics such as normalized root mean square error (NRMSE) were used to assess the signal prediction fidelity. In addition, the fidelity of the estimated dictionary weights was assessed using normalized Wasserstein distance (NWD) metric, evaluating how the weight vector is affected by the process of leaving out measurements from a given viscosity level.
\\
\\
\noindent {\bf Results:} 
Validation results on synthetic signals exhibit high fidelity weight vector and transfer function estimations even at a very low SNR level of 1. 
Experiments on in-house MPS setup demonstrate that the proposed algorithm can successfully predict the MNP signals at untested viscosities with NRMSE values 
remaining below 1.51\% and 3.5\% for the two tested MNPs, at all viscosity levels and DF settings. Importantly, the predicted signals successfully capture the trends in the MNP signal as a function of viscosity. The quantitative assessments using the NWD metric indicate a robustness in dictionary weight estimation, with NWD remaining at low values below 0.10 and 0.07 for the two tested MNPs.
\\
\\
\noindent {\bf Conclusions:} 
The proposed calibration-free iterative dictionary-based algorithm jointly estimates the dictionary weights and non-model based dynamics, successfully predicting the MNP signal and capturing the trends as a function of viscosity. The proposed approach can be utilized for the optimization of DF settings or MNP type for applications such as viscosity mapping or temperature mapping using MPI.\\
\\
\noindent Key words: magnetic particle imaging, signal prediction, magnetic nanoparticles, coupled Brown-Neel rotation model, model-based dictionary.

\end{abstract}


\setlength{\baselineskip}{0.7cm} 

\pagenumbering{arabic}
\setcounter{page}{1}
\pagestyle{fancy}

\section{Introduction}
Magnetic particle imaging (MPI) allows quantification and spatial mapping of magnetic nanoparticle (MNP) distributions with high resolution and sensitivity~\cite{Gleich2005,Saritas2013}.
MPI has a variety of applications including stem cell tracking~\cite{Zheng2015,Sehl2020}, angiography~\cite{Mohtashamdolatshahi2020}, cancer imaging \cite{Tay2021}, localized hyperthermia~\cite{Chandrasekharan2020}, interventional imaging~\cite{Bakenecker2018,Wegner2022}, inflammation imaging~\cite{Chandrasekharan2021}, and viscosity and/or temperature mapping~\cite{Utkur2019,Utkur2022}. 
MPI utilizes an oscillating drive field (DF) to generate a time-varying magnetization response for MNPs inside a field free region (FFR), which in turn induces a signal on a receive coil.
Two main mechanisms govern the magnetization dynamics of MNPs: Néel and Brownian processes~\cite{Shasha2020}. 
In the Néel process, magnetic moment of MNP internally aligns with the applied field without any physical rotation. This process is affected by magnetic core diameter and magnetic anisotropy originating from the shape and crystal structure of the MNP, as well as temperature.
In the Brownian process, MNP physically rotates to align its magnetic moment with the applied field. This process is affected by core diameter and hydrodynamic diameter of the MNP, as well as viscosity and temperature.
Brownian and Néel processes occur in a coupled fashion, with magnetic moment and particle rotations affecting each other. In addition, interparticle magnetodipolar interactions also affect the overall magnetization behavior~\cite{Wu2019,Moor2022}.
For applications such as temperature and viscosity mapping, the optimal DF settings that maximize sensitivity to these environmental conditions depend highly on MNP type and need to be determined via extensive experiments ~\cite{Utkur2022}. A better understanding of magnetization dynamics can enable optimization of DF settings and/or MNP type for a given application. 
Various theoretical models describing magnetization dynamics have been proposed with the goal of explaining the experimental results in MPI. 
One approach is to decouple the Brownian and Néel processes by inhibiting or neglecting one of these processes, such as immobilizing MNPs to inhibit the Brownian process~\cite{Albers2022,Albers2022_Feb}, neglecting the Brownian rotation once the easy axis is aligned with the field direction~\cite{Weizenecker2012}, ignoring Néel rotation under low viscosity and strong anisotropy~\cite{Martens2013}, or ignoring Brownian rotation due to the dominant Néel rotation observed in simulations~\cite{Graeser2015}. 
Recently, the Néel rotation model was utilized to create model-based dictionaries containing simulated signals for MNPs with different magnetic properties. These dictionaries were then employed for magnetic parameter estimation via data fitting to magnetic particle spectrometer (MPS) measurements~\cite{Albers2022} or system matrix measurements~\cite{Albers2022_Feb}, with the goal of achieving model-based system calibration. 
However, models that consider the interdependence between the Néel and Brownian rotations can provide a more comprehensive understanding of magnetization dynamics. Stochastic Langevin equations describing this coupling were solved in previous simulation studies \cite{Reeves2015,Shasha2017,Engelmann2019}. Another study presented an important step in this direction by deriving coupled differential equations for magnetic moment and particle rotations and transforming them into a system of ordinary differential equations (ODE) where a numerical solution is possible~\cite{Weizenecker2018}. 

In this study, we propose a calibration-free iterative dictionary-based algorithm that uses the coupled Brown–Néel rotation model, with the goal of predicting the MNP signal. The proposed algorithm jointly estimates dictionary weights and non-model based dynamics based on available measurements, and utilizes this information to predict the MNP signal at untested viscosity levels.
This approach accounts for model based dynamics, as well as the system and magnetization dynamics that the model can not explain. 
Using MPS experiments performed in the biologically relevant viscosity range of 0.89 to 15.33~mPa$\cdot$s and DF frequencies between 0.25 to 2~kHz, we show that the proposed approach can successfully predict the MNP signals at untested viscosities.

\section{Theory}

\subsection{Coupled Brown-Néel Rotation Model}

\begin{figure}[!t]
   \begin{center}
   \includegraphics{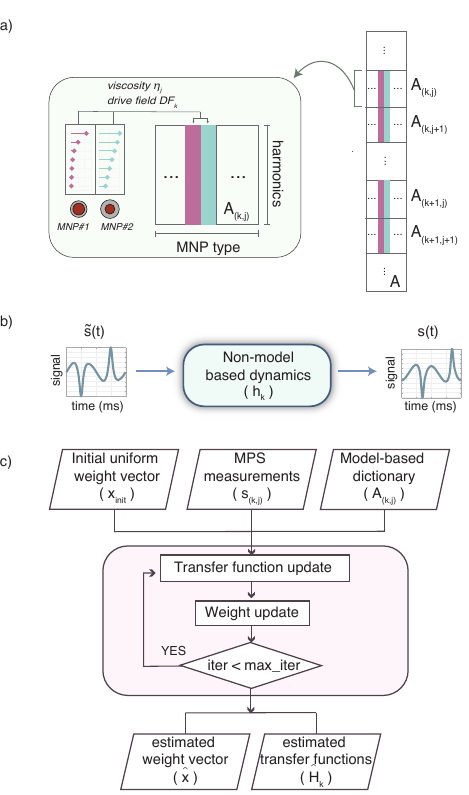} 
   \captionv{12}{}{Simulated MNP signals comparing coupled Brown–Néel rotation model and other models for two different viscosity levels of $\eta$~=~0.89~mPa$\cdot$s and 15.33~mPa$\cdot$s at two different DF settings: (a) at 10~mT, 250~Hz, and (b) at 10~mT, 1~kHz. The MNP signals were calculated by taking the time derivative of the simulated magnetization responses and then filtering out the fundamental harmonic. The magnetic parameters of the simulated MNP were $d_c$ = 20~nm core diameter, $d_h$ = 60~nm hydrodynamic diameter, and $K$ = 6~kJ/m$^{3}$ magnetic anisotropy constant. \label{fig:model_comparison}
    }  
    \end{center}
\end{figure}

To simulate the magnetization dynamics of MNPs, we utilized the coupled Brown–Néel rotation model~\cite{Weizenecker2018}, which assumes single-domain, noninteracting, uniaxially symmetric particles with uniaxial magnetic anisotropy.
Fig.~\ref{fig:model_comparison} compares the simulated MNP signals from the coupled model with those from computationally simpler magnetization models, at two different viscosity levels and two different DF settings. The MNP parameters were chosen based on the literature~\cite{Eberbeck2013}.
The Langevin model assumes that the relaxation effects are negligible and the particles are at equilibrium. The Néel rotation results are shown for two different scenarios: (1) the aligned Néel model, where the easy axis (i.e., the easiest direction to magnetize the material) and applied field were assumed to be aligned~\cite{Deissler2013}, and (2) the Brown–Néel model at extremely high viscosity to block Brown rotation, where the projection of the easy axis onto the applied field direction was assumed to have a zero mean~\cite{Weizenecker2018}. These two Néel models yield similar relaxation delays, albeit with different signal amplitudes due to the differences in the initial alignment of the easy axis. 
Because the Langevin model and the two Néel models do not incorporate effects of viscosity on magnetization, their signals remain unchanged with viscosity. While the pure Brown model incorporates viscosity effects, its signal shows a low amplitude and unrealistically large relaxation delay, especially at higher frequency. In contrast, the coupled Brown–Néel model shows reduced relaxation delay when compared to both the Néel and Brown models, while capturing the effects of viscosity.  Consequently, the coupled Brown–Néel model offers a distinct advantage by more comprehensively capturing the magnetization dynamics.

\subsection{Dictionary Formation}

In this work, the MNP signals simulated using the coupled Brown–Néel model were used to construct a dictionary matrix, with each column containing the odd harmonics of the Fourier transform of the simulated signal for MNPs with different magnetic parameters. Because the signal changes with DF settings and viscosity, we built separate dictionaries for each case. Namely, $\mathbf{A_{\left( k,j\right)}}\in\mathbb{C}^{M_k \times N}$ is the dictionary matrix constructed in the frequency domain for the $k^{th}$ DF setting DF$_k$ and the $j^{th}$ viscosity level $\eta_j$, where ${\mathbb{C}}$ denotes the set of complex numbers. Here, $N$ is the number of simulated particles, and $M_k$ is the number of selected harmonics for DF$_k$. 
The constructed dictionaries can then be utilized to express the measured MNP signal as a weighted linear combination of the dictionary columns. 

\subsection{Incorporating Non-Model Based Dynamics}

Solving a linear system of equations with model-based dictionaries and measured MNP signals enables the weight estimation of dictionary elements. However, solving directly for the dictionary weights is error-prone due to the presence of non-model based dynamics: First, the system response of the measurement setup alters the measured signals. If this effect is not accounted for, the estimated dictionary weights will have to compensate for the system response. However, this compensation may not work well for measurements performed at multiple frequencies. Second, the employed magnetization model may not fully account for real-life MNP dynamics, causing the dictionary weights to include arbitrary particles to explain the measured MNP dynamics.

\begin{figure}[!t]
   \begin{center}
   \includegraphics{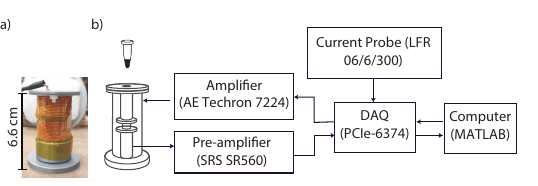} 
   \captionv{12}{}{Schematic description of the proposed algorithm. (a) Given the spanned signal ($\tilde{s}(t)$) as input, the measured signal ($s(t)$) is modeled as the output of an LTI system that represents the non-model based dynamics at drive field DF$_k$. (b) The proposed algorithm jointly estimates the weight vector and the transfer functions, given a  uniform initial weight vector, MPS measurements, and the model-based dictionary as input. The algorithm alternates between the transfer function update and weight update steps for a predetermined maximum number of iterations. \label{fig_main}
    }  
    \end{center}
\end{figure}

We propose incorporating these non-model based dynamics into the problem via a linear and time-invariant (LTI) system approach, as shown in Fig.~\ref{fig_main}(a). Accordingly, the measured MNP signal in frequency domain is expressed as the multiplication of the MNP signal spanned by the dictionary and the transfer function of an LTI system:
\begin{linenomath*}
 \begin{align} \label{eq3}
H_{k}(f) \hspace{0.15em} \tilde{S}_{k,j}(f)  &= S_{k,j}(f)
\end{align}
\end{linenomath*}
Here, $H_{k}$ is the transfer function of the LTI system representing non-model based dynamics at DF$_k$. $S_{k,j}$ and $\tilde{S}_{k,j}$ are the measured signal and the signal spanned by the dictonary at DF$_k$ and $\eta_j$, respectively.
We assume that $H_{k}$ does not depend on the sample viscosity but depends on the DF settings (i.e., DF frequency and amplitude), as both the impedances in the measurement setup and the dipolar interactions depend on these parameters~\cite{Berkov2006,Ota2015}.
Additionally, $H_{k}$ depends on the tested MNP sample due to the differences in dipolar interactions and magnetic anisotropy types. The dipolar interactions are affected by whether the MNPs are multi-core or single-core~\cite{Ludwig2017,Ahrentorp2015}, as well factors such as interparticle distances~\cite{Wu2019,Moor2022}. Besides dipolar interaction, the magnetic anisotropy type varies for different MNP samples~\cite{Zhao2020}.

\subsection{Joint Estimation of Dictionary Weights and Transfer Functions}
We propose finding the dictionary weights that, together with the transfer functions, explain the measured MNP signals. 
Accordingly, the problem is formulated as follows:
\begin{linenomath*}
\begin{align}\label{eq:overall_problem}
\begin{split}    \min_{\mathbf{x}\in\mathbb{R}^N,\{\mathbf{H_{k}}: \forall k \in D\}} \sum_{k\in D} 
   \sum_{j\in V}
  &\left\|\mathbf{H_{k}} \mathbf{A_{\left( k,j\right)}}   \mathbf{x}  - \mathbf{s_ {\left(k,j\right)}} \right\|_{2}^2  \\
 &\textrm{s.t.} \quad \mathbf{x} \geq 0 
\end{split}
\end{align}
\end{linenomath*}
Here, $\mathbf{s_ {\left(k,j\right)}}\in\mathbb{C}^{M_k \times 1}$ is the measurement vector at DF$_k$ and $\eta_j$, $\mathbf{H_{k}}\in\mathbb{C}^{M_k \times M_k}$ is a diagonal matrix with the diagonal entries containing the transfer function values at the harmonics, and $\mathbf{x}\in\mathbb{R}^{N\times1}$ is the non-negative dictionary weight vector to be estimated, where $\mathbb{R}$ denotes the set of real number. Lastly, $D$ is the set of all DF indices, and $V$ is the set of all viscosity indices.

This problem can be solved by adopting an alternating minimization approach in two steps: \textit{transfer function update} and \textit{weight update}. The proposed algorithm, summarized in Fig.~\ref{fig_main}(b), is initialized with a normalized uniform weight vector. Then, the algorithm alternately optimizes either the weight vector or the transfer functions while keeping the other fixed, for a predetermined maximum number of iterations. The two optimization steps are described in detail below.

\textbf{Transfer Function Update:} Keeping the weight vector $\mathbf{x}$ fixed, the optimization problem in Eq.~\ref{eq:overall_problem} becomes
\begin{linenomath*}
\begin{align}
    \min_{\{\mathbf{H_{{k}}}: \forall k \in D\}}\sum_{k\in D}
    \sum_{j\in V}
   \left\|\mathbf{H_{k}} \hspace{0.15em} \mathbf{\tilde{s}_ {\left(k,j\right)}} - \mathbf{s_ {\left(k,j\right)}} \right\|_{2}^2
\end{align}
\end{linenomath*}
\noindent where 
\begin{linenomath*}
\begin{align}\label{eq:undistortedSignal}
\mathbf{\tilde{s}_ {\left(k,j\right)}} = \mathbf{A_ {\left(k,j\right)}}   \mathbf{x} 
\end{align}
\end{linenomath*}
\noindent Here, $\mathbf{\tilde{s}_ {\left(k,j\right)}}\in\mathbb{C}^{M_k \times 1}$ is fixed, since $\mathbf{x}$ is fixed. This problem is separable in DF indices. Hence, the optimization for each DF$_k$ can be performed independently from others as
\begin{linenomath*}
\begin{align}
    \mathbf{\hat{H}_{k}} = \argminB_{\mathbf{H_{{k}}}\in\mathbb{C}^{M_k \times M_k}} 
    \sum_{j\in V}
   \left\| \mathbf{H_{k}} \hspace{0.15em} \mathbf{\tilde{s}_ {\left(k,j\right)}} - \mathbf{s_ {\left(k,j\right)}} \right\|_{2}^2  
\end{align}
\end{linenomath*}
\noindent where $\mathbf{\hat{H}_{k}}\in\mathbb{C}^{M_k \times M_k}$ is the estimated transfer function. Since $\mathbf{H_{k}}$ is a diagonal matrix, this problem is also separable in harmonic indices. 
The problem for the $m^{th}$ harmonic is
\begin{linenomath*}
\begin{align}
\begin{split}
    \min_{H_{k}(mf_k)} 
    \sum_{j\in V}
   \left\|H_{k}(mf_k) \hspace{0.15em} \tilde{S}_{k,j}(mf_k) - S_{k,j}(mf_k) \right\|_{2}^2  
   \end{split}
\end{align}
\end{linenomath*}
\noindent where $f_k$ is the fundamental frequency for DF$_k$. Using least square solution via Moore-Penrose pseudoinverse gives
\begin{linenomath*}
\begin{align}\label{Hupdate}
\hat{H}_{k}(mf_k) =    \frac{\sum\limits_{j\in V} \tilde{S}_{k,j}^*(mf_k) \hspace{0.15em} S_{k,j}(mf_k)}{\sum\limits_{j\in V} \left(\tilde{S}_{k,j}(mf_k)\right)^2}
\end{align}
\end{linenomath*}
\noindent where $\hat{H}_{k}(mf_k)$ is a scalar and superscript $*$ denotes complex conjugation. After solving Eq.~\ref{Hupdate} for each DF setting DF$_k$ and $m$, the transfer function update step is completed.

\textbf{Weight Update:} Keeping the transfer functions $\mathbf{{H}_{k}}$ fixed, the optimization problem in Eq.~\ref{eq:overall_problem} becomes a standard nonnegative least square (NNLS) problem as follows:
\begin{linenomath*}
\begin{align}\label{weight_up}
\begin{split}
    \mathbf{\hat{x}} = \argminB_{\mathbf{x}\in\mathbb{R}^N}\sum_{k\in D}
    \sum_{j\in V}
   &\left\| \mathbf{{H}_{k}} \mathbf{A_ {\left(k,j\right)}}   \mathbf{x}  - \mathbf{s_ {\left(k,j\right)}} \right\|_{2}^2  \\&\textrm{s.t.} \quad \mathbf{x} \geq 0 
\end{split}   
\end{align}
\end{linenomath*}
This problem can be solved by vertically concatenating the matrices $\mathbf{{H}_{k}} \mathbf{A_ {\left(k,j\right)}}$ for different DF$_k$ and $\eta_j$, with the real and imaginary parts separated and concatenated. Then, the problem can be expressed as:
\begin{linenomath*}
\begin{align}\label{eq:weight_up_conc}
\begin{split}
    \mathbf{\hat{x}} = \argminB_{\mathbf{x}\in\mathbb{R}^N}
   &\left\| \mathbf{\Bar{A}}   \mathbf{x}  - \mathbf{\Bar{s}} \right\|_{2}^2  \\&\textrm{s.t.} \quad \mathbf{x} \geq 0 
\end{split}   
\end{align}
\end{linenomath*}
Here, $\mathbf{\Bar{A}} \in\mathbb{R}^{2 J M \times N}$ is the concatenated matrix and $\mathbf{\Bar{s}} \in\mathbb{R}^{2 J M \times 1}$ is the concatenated measurement vector, where $M = \sum_{k\in D}M_k$ is the total number of selected harmonics across all DF settings and $J$ is the cardinality of set $V$ (i.e., the number of viscosities considered). 

It should be noted that the problem in Eq.~\ref{eq:overall_problem} has infinitely many solution pairs $(\hat{\mathbf{x}},\hat{\mathbf{H}}_k)$, due to the fact that an arbitrary scaling given as $(\alpha\hat{\mathbf{x}},\hat{\mathbf{H}}_k/\alpha)$ for some constant $\alpha$ yields the same minimization error. However, such a scaling does not affect signal fitting or signal prediction stages, where $\hat{\mathbf{x}}$ and $\hat{\mathbf{H}}_k$  are multiplied to cancel out $\alpha$, as described below.

\subsection{MNP Signal Fitting}\label{sec:signalFitting}
We define the \textit{spanned signal} as the signal spanned by the dictionary alone (i.e., excluding the non-model based dynamics) given the estimated weight vector $\mathbf{\hat{x}}$, i.e.,
\begin{linenomath*}
\begin{align}\label{eq:signal_spanned}
\mathbf{\tilde{s}_ {\left(k,j\right)}} = \mathbf{A_ {\left(k,j\right)}}   \hat{\mathbf{x} }
\end{align}
\end{linenomath*}
\noindent Next, we define the final \textit{fitted signal} as the signal after incorporating the estimated transfer functions $\mathbf{\hat{H}_{k}}$, i.e.,
\begin{linenomath*}
\begin{align}\label{eq:signal_fitted}
\begin{split}
\mathbf{\hat{s}_ {\left(k,j\right)}} = 
\mathbf{\hat{H}_{k}} \mathbf{\tilde{s}_ {\left(k,j\right)}} = \mathbf{\hat{H}_{k}} \mathbf{A_ {\left(k,j\right)}}   \hat{\mathbf{x} }
\end{split}   
\end{align}
\end{linenomath*}
Here, $\mathbf{\tilde{s}_ {\left(k,j\right)}},\mathbf{\hat{s}_ {\left(k,j\right)}}\in\mathbb{C}^{M_k \times 1}$ contain the harmonics of the spanned and fitted signals, respectively.  

\subsection{MNP Signal Prediction}\label{sec:signal_prediction}
For predicting the signal at an untested viscosity level (i.e., a viscosity level not included when estimating $(\hat{\mathbf{x}},\hat{\mathbf{H}}_k)$), we propose constructing a new dictionary at the untested viscosity level using the coupled Brown-Néel model. Then, the signal spanned by the new dictionary can be calculated as:
\begin{linenomath*}
\begin{align}\label{eq:signal_spanned_predicted}
\begin{split}
\mathbf{\tilde{s}_ {\left(k,e\right)}} = \mathbf{A_ {\left(k,e\right)}} \mathbf{\hat{x}} 
\end{split}   
\end{align}
\end{linenomath*}
Incorporating the transfer functions, the final \textit{predicted signal} at the untested viscosity level can be expressed as: 
\begin{linenomath*}
\begin{align}\label{eq:signal_predicted}
\begin{split}
\mathbf{\hat{s}_ {\left(k,e\right)}} 
=\mathbf{\hat{H}_{k}} \mathbf{\tilde{s}_ {\left(k,e\right)}}
=\mathbf{\hat{H}_{k}} \mathbf{A_ {\left(k,e\right)}} \mathbf{\hat{x}}
\end{split}   
\end{align}
\end{linenomath*}
Here, $\mathbf{A_ {\left(k,e\right)}}\in\mathbb{C}^{M_k \times N}$ is the new dictionary at DF$_k$ at the untested viscosity, where the index `$\mathbf{e}$' indicates that this viscosity level was \textit{excluded} from consideration when estimating $(\hat{\mathbf{x}},\hat{\mathbf{H}}_k)$. Lastly, $\mathbf{\tilde{s}_ {\left(k,e\right)}} , \mathbf{\hat{s}_ {\left(k,e\right)}} \in\mathbb{C}^{M_k \times 1}$ contain the harmonics of the spanned and predicted signals at the untested viscosity level, respectively.

\section{Methods}

\subsection{Magnetization Simulations for Dictionary Formation}
Magnetization simulations were performed using the coupled Brown–Néel rotation model~\cite{Weizenecker2018} for MNPs with different magnetic core diameters ($d_c$), surface coating diameters ($d_s$), and uniaxial magnetic anisotropy constants ($K$). Table~\ref{tab:MagProperties_united} lists the ranges for these MNP parameters, which were chosen to cover the typical values reported in the literature~\cite{Eberbeck2013,Lanier2019}. The hydrodynamic diameter ($d_h$) was assumed to be the sum of the surface coating diameter and magnetic core diameter (i.e., $d_h = d_s+d_c$). Saturation magnetization ($M_s$) and temperature were assumed to be 360 kA/m~\cite{Eberbeck2013}, and 298.5~K, respectively. Gilbert damping constant and gyromagnetic ratio were taken as $0.1$ and $1.75\times 10^{11}$~rad/(s$\cdot$T), respectively~\cite{Deissler2013}.

\begin{table}[h]
\begin{center}
\captionv{10}{}{Parameters for Magnetization Simulations
\label{tab:MagProperties_united}
\vspace*{2ex}
}

\begin{tabular}{|c|c|c|}
  \hline 
  \multirow{3}{*}{\textbf{MNP Parameters}} & $d_c\hspace{0.3em} \text{(nm)}$    & $\{5,6,7,....,30 \}$ \\
  \cline{2-3}  & $d_s \hspace{0.3em}\text{(nm)}$ & $\{15,20,25,....,100 \}$ \\ 
  \cline{2-3} & $K \hspace{0.3em} (\text{kJ/m}^{3})$ & $\{1,2,3,....,10 \} $\\ 
   \hline 
  \multirow{2}{*}{\textbf{DF Settings}} & $f_d \hspace{0.3em} \text{(Hz)}$&$\{250,1000,2000\}$ \\
  \cline{2-3}& $B_p \hspace{0.3em} \text{(mT)}$ & $\{10,15\}$    \\
  \hline
  \multirow{1}{*}{\textbf{Viscosity}} & $\eta \hspace{0.3em}$ (mPa$\cdot$s) & \{0.89, 1.45, 2.08, 3.32, 8.31, 15.33\} \\
  \hline
\end{tabular}
\end{center}
\end{table}

At a given DF$_k$ and $\eta_j$, the magnetization response for a given set of magnetic parameters was simulated by solving the ODEs for the coupled Brown–Néel rotation model~\cite{Weizenecker2018}, via the built-in variable-step variable-order solver \textit{ode15s} of MATLAB. As initial condition, the particles were assigned to have zero average magnetization. The magnetization response was simulated for three DF periods. The first two periods were discarded to avoid transient states, and the remaining period was sampled at a rate of 2 MS/s. Next, the corresponding MNP signal was computed by taking the numerical time derivative of the magnetization response, multiplied by the respective core volumes~\cite{Kluth2018}. Finally, the selected harmonics in Fourier domain were inserted into the corresponding column of the dictionary matrix $\mathbf{A_{\left( k,j\right)}}$.

Simulations were performed on a server with Intel Xeon E5-2667. In total, 4,680 different MNPs were simulated at 6 different DF settings and 6 different viscosity levels, resulting in 168,480 distinct simulations (see Table~\ref{tab:MagProperties_united}).  The DF settings consisted of two different DF amplitudes ($B_p$) of 10~mT and 15~mT, and three different DF frequencies ($f_d$) of 250~Hz, 1~kHz, and 2~kHz. These relatively low DF frequencies were selected as they were previously shown to have a high sensitivity to changes in viscosity~\cite{Utkur2022,Utkur2019}.  The viscosity levels ranged between 0.89-15.33~mPa$\cdot$s.

\subsection{Implementation of the Proposed Algorithm}
The maximum number of iterations for the proposed algorithm (see Fig.~\ref{fig_main}(b)) was chosen heuristically as 2000 to ensure convergence. For the weight update stage, the built-in function \textit{lsqnonneg} of MATLAB was utilized to solve the NNLS problem given in Eq.~\ref{eq:weight_up_conc}, with the termination tolerance on $\mathbf{x}$ set to 
$10^{-9} \times \left\| \mathbf{\Bar{A}} \right\|_1 \times \left\| \mathbf{\Bar{s}}\right\|_1$, where $\left\| {\cdot}\right\|_1$ denotes $\ell_1$-norm. 
Note that the tolerance changes at each iteration as $\mathbf{\Bar{A}}$ changes during the transfer function update step.

\subsection{Validation on Synthetic MNP Signals}
For the case of experimental measurements, ground-truth weight vectors and transfer functions are generally not available. Therefore, to validate the performance of the proposed algorithm, we first generated synthetic signals for a known set of MNP parameters for the same DF settings and viscosities listed in Table~\ref{tab:MagProperties_united}, and then used the proposed algorithm to estimate the underlying weight vector and transfer functions. 
Ideal transfer functions with unit magnitude and zero phase response were assumed.
Zero mean white Gaussian noise, with identical standard deviation (SD) for all DF settings,  was added to the synthetic signals in time domain. Two different signal-to-noise ratio (SNR) levels of 1 and 10 were considered, mimicking the range of SNRs in MPS experiments. SNR was defined as the ratio of the noise SD to the maximum synthetic signal amplitude at 0.89 mPa$\cdot$s for 10 mT and 250 Hz. 

\begin{figure}[!t]
   \begin{center}
   \includegraphics{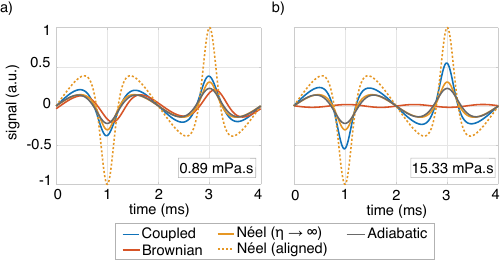} 
   \captionv{12}{}{In-house arbitrary waveform magnetic particle spectrometer (MPS) setup. (a) A picture of the MPS setup, featuring a low-inductance drive coil and a position-adjustable gradiometric receive coil.  (b) The signal workflow of the transmit/receive chain is depicted with arrows.\label{setup_withPhoto}
    }  
    \end{center}
\end{figure}

\subsection{MPS Experiments}
MPS experiments were performed on two different MNPs, at the DF settings and viscosity levels listed in Table~\ref{tab:MagProperties_united}.

\textbf{MPS Setup:} The in-house arbitrary waveform MPS setup (see Fig.~\ref{setup_withPhoto}) enabled arbitrary choice of DF frequencies without the need for impedance matching \cite{Tay2016}. The drive coil had a relatively low inductance of 13.3~$\mu$H, and had 0.66~mT/A sensitivity with 95\% homogeneity in a 0.7~cm-long region in the measurement chamber, measured using a Hall effect gaussmeter (LakeShore 475 DSP Gaussmeter). The receive coil had a two-section gradiometric design, enabling direct feedthrough suppression via manual adjustment. During data acquisition, the DF waveform designed on MATLAB was sent to the power amplifier (AE Techron 7224) via a data acquisition card (NI USB-6383), and the amplified waveform generated the desired DF within the drive coil. A Rogowski current probe (LFR 06/6/300, PEM Ltd.) was utilized for calibration and real-time monitoring of the DF waveform. The received MNP signal was amplified and band-pass filtered with cutoffs at 100~Hz and 1~MHz using a low-noise preamplifier (SRS SR560), and sampled at a rate of 2~MS/s.

\textbf{Sample Preparation:} Experiments were performed on two different commercially available MNPs: Vivotrax (Magnetic Insight Inc., USA) and Perimag (Micromod GmbH, Germany), with 5.5 and 8.5 mg/ml undiluted Fe concentrations, respectively. For each MNP, samples at 6 different viscosity levels between 0.89-15.33~mPa$\cdot$s were prepared. Each sample contained 26.2~\textmu l undiluted MNP solution, with glycerol and deionized water added to reach the targeted viscosity level at a total volume of 70~\textmu l. Table~\ref{tab:glyDI} lists the viscosity levels and the corresponding glycerol volume percentages at 25$^o$C~\cite{Cheng2008}. 

\begin{table}[h]
\begin{center}
\captionv{10}{}{Glycerol volume percentages of samples at 25$^o$C
\label{tab:glyDI}
\vspace*{2ex}
}

\begin{tabular}{|p{105pt}|p{25pt}|p{25pt}|p{25pt}|p{25pt}|p{25pt}|p{25pt}|}
\hline
Viscosity (mPa$\cdot$s)& 0.89 & 1.45 &2.08 & 3.32 &8.31 & 15.33
\\ \hline
Glycerol vol. ($\%$) & 0 & 15.1 &24.8 & 35.8&53.2 & 62.5 \\
\hline
\end{tabular}
\end{center}
\end{table}

\textbf{Experiments:} At each DF setting and viscosity level, 3 repeated measurements were performed. Each repetition consisted of 8 averaged pulse acquisitions,  with each pulse containing 20 DF periods.
Before and after each MNP signal measurement, empty chamber measurements spanning all DF settings were performed for baseline subtraction. The fundamental harmonic was filtered out, followed by harmonic selection. The number of selected harmonics, $M_k$, for each DF setting DF$_k$ was determined with a signal-to-background ratio (SBR) thresholding, where SBR was defined as the ratio of the MNP signal amplitude to the baseline signal amplitude at a given harmonic. The SBR threshold was set as 15, resulting in $M_k$ ranging between 3-33 across different DF settings.

\textbf{MNP Signal Fitting:} Using MPS measurements at all 6 DF settings and 6 viscosity levels,
the proposed algorithm was utilized to estimate the weight vectors and transfer functions for each MNP, separately. Then, the fitted signals were computed using Eq.~\ref{eq:signal_fitted}. 

\textbf{MNP Signal Prediction:} MPS measurements from one of the viscosity levels were left out entirely, and the weight vector and transfer functions were estimated with the remaining measurements from other viscosity levels for each MNP, separately. The signals at the left-out viscosity level were then predicted using Eq.~\ref{eq:signal_predicted}. These steps were repeated by excluding each viscosity level one by one.

\subsection{Quantitative Evaluation Metrics}\label{sec:evaluation_metrics}
\textbf{Normalized Root Mean Square Error (NRMSE):} To evaluate the overall quality of fitted/predicted signals, normalized root mean square error (NRMSE) was computed as follows:
\begin{linenomath*}
\begin{align}
\text{NRMSE}(\mathbf{v},\mathbf{\hat{v}}) = \frac{\left\|\mathbf{v}-\mathbf{\hat{v}}\right\|_2}{\sqrt{n}\left(\max (\mathbf{v})  - \min (\mathbf{v})\right)}
\end{align}
\end{linenomath*}
Here, $\mathbf{v}$ and $\mathbf{\hat{v}}$ are the measured and fitted/predicted signal vectors in time domain, respectively. In addition, $n$ is the length of $\mathbf{v}$, and $\max (\mathbf{v}) $ and $\min (\mathbf{v})$ denote the maximum and minimum values in the vector $\mathbf{v}$, respectively.

\textbf{Relaxation Time Constant ($\tau$):} To assess the similarity of the relaxation dynamics between the measured and fitted/predicted signals, relaxation time constants ($\tau$) were calculated using TAURUS (TAU estimation via Recovery of Underlying mirror Symmetry)~\cite{Utkur2022,Arslan2022}. These $\tau$ values were normalized by the DF period and converted to percentages as
\begin{linenomath*}
\begin{align}
\hat{\tau} &= \frac{\tau}{T_d} \cdot 100
\end{align}
\end{linenomath*}
\noindent where $T_d = 1/f_d$ is the DF period. This normalization enables comparison of $\tau$ values at different DF frequencies.

\textbf{Zero-crossing Time (t$_{\textit{zc}}$):} To evaluate the similarity of the signal widths between the measured and fitted/predicted signals, we computed zero-crossing time ($t_{zc}$), which we define as the time interval between two zero-crossing points around a signal peak in one-half cycle signal. These $t_{zc}$ values were normalized by the DF period and converted to percentages as
\begin{linenomath*}
\begin{align}
\hat{t}_{zc}  &= \frac{{t}_{zc}}{T_d} \cdot 100
\end{align}
\end{linenomath*}

\textbf{Signal Root Mean Square (v${_\textit{rms}}$):} To compare the signal amplitudes, root mean square amplitudes of the signals ($v_{rms}$) were calculated and normalized as follows:
\begin{linenomath*}
\begin{align}
\hat{\text{v}}_{rms}  &= \frac{\|\mathbf{v}\|_2}{\sqrt{n}}\cdot\frac{1}{ f_d \cdot B_p\cdot m_{Fe}}
\end{align}
\end{linenomath*}
\noindent Note that this computation factors out the expected scalings of the signal with DF frequency and amplitude, so that the effects of relaxation on signal amplitude can be delineated. In addition, normalization by the iron amount in the sample ($m_{Fe} (\text{g})$) enables comparison of relative efficiencies of different MNPs (i.e., Perimag and Vivotrax in this case).

\textbf{Normalized Wasserstein Distance:} To evaluate the similarity between two weight vectors, normalized Wasserstein distance (NWD) was utilized \cite{Wei2022}. This metric, also known as earth mover's distance, computes the cost of turning one probability mass function (PMF) into the other via optimal mass transport. For this evaluation, the weight vectors were converted to PMFs by normalizing them to have unit sum. Then, NWD was calculated as follows:
\begin{linenomath*}
\begin{align}
NWD (\mathbf{P_1},\mathbf{P_2}) = \frac{1}{L-1} \|\mathbf{F_1-F_2}\|_1
\end{align}
\end{linenomath*}
Here, $\mathbf{P_1},\mathbf{P_2}\in\mathbb{R}^{L \times 1}$ are the compared PMF vectors, $\mathbf{F_1},\mathbf{F_2}\in\mathbb{R}^{L \times 1}$ are the corresponding cumulative distribution function (CDF) vectors, and $L$ is the vector length. NWD ranges between 0 and 1, where 0 indicates that the two PMFs are identical and 1 indicates that they are maximally different.

\section{Results}

\begin{figure}[!t]
   \begin{center}
   \includegraphics{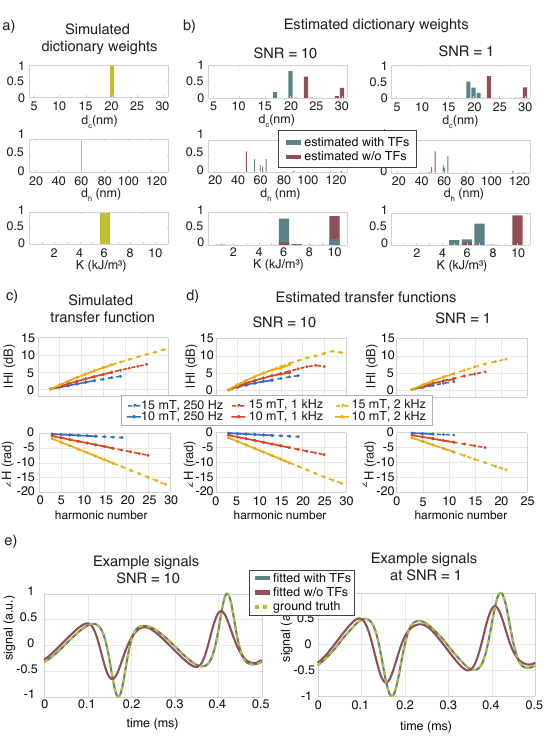} 
   \captionv{12}{}{Validation results
    on synthetic MNP signals generated for a single particle with $d_c$~=~20~nm, $d_h$~=~60~nm, and $K$~=~6~kJ/m$^3$, assuming unit magnitude and zero phase transfer functions. Estimated weight vectors at (a) SNR = 10 and (d) SNR = 1, displayed as PMFs for $d_c$, $d_h$, and $K$. Estimated magnitude and phase responses for the transfer functions for 6 different DF settings at (b) SNR = 10 and (e) SNR = 1. Example synthetic, fitted, and spanned signals for (c) SNR = 10 and (f) SNR = 1 for $\eta$ = 0.89 mPa$\cdot$s at 10 mT, 1 kHz DF setting. 
    The estimated transfer functions were normalized together, such that the 3$^{\textrm{rd}}$ harmonic for 10~mT, 250~Hz DF setting has 0~dB magnitude response.\label{fig:synt_results}
    }  
    \end{center}
\end{figure}

\subsection{Validation Results on Synthetic MNP signals}

Synthetic MNP signals were generated for a single particle with $d_c$ = 20~nm, $d_h$ = 60~nm, and $K$ = 6~kJ/m$^3$ (i.e., the same parameters as in Fig.~\ref{fig:model_comparison}), at the same 6 DF settings and 6 viscosity levels listed in Table~\ref{tab:MagProperties_united}, assuming unit magnitude and zero phase transfer functions. Then, the noise-added synthetic MNP signals were fed to the proposed algorithm. Figure~\ref{fig:synt_results} shows the estimated weight vectors, transfer functions, and fitted signals for the cases of SNR = 10 and SNR = 1. In Fig.~\ref{fig:synt_results}(a) and \ref{fig:synt_results}(d), the estimated weight vectors are displayed as marginal PMFs for $d_c$, $d_h$, and $K$. At SNR = 10, the proposed algorithm almost exactly recovers the ground-truth weight vector, whereas there are slight deviations at SNR = 1. The NWDs between the estimated and the ground-truth weight vectors are 0.0007, 0.0009, and 0.0023 for the marginal PMFs of $d_c$, $d_h$, and $K$ at SNR = 10. At SNR = 1, the respective NWDs are 0.0088, 0.0159, and 0.0246. These results demonstrate high fidelity estimations even at low SNR.

Figures~\ref{fig:synt_results}(b) and \ref{fig:synt_results}(e) show the transfer functions estimated jointly with the weight vectors at SNR = 10 and SNR = 1, respectively. The magnitude responses remain very close to 1 (0 dB) at both SNR levels, increasing slightly at higher harmonics to a maximum of 1.01 (0.098 dB) and  1.03 (0.24 dB) at SNR = 10 and SNR = 1, respectively. Similarly, the phase responses remain very close to zero, increasing to a maximum of 0.006 rad and 0.022 rad at higher harmonics at SNR = 10 and SNR = 1, respectively. These slight deviations are potentially due to reduced  SBR at higher harmonics.

Lastly, Figs.~\ref{fig:synt_results}(c) and \ref{fig:synt_results}(f) show example syhthetic, fitted and spanned signals at DF settings of 10~mT and 1~kHz, at $\eta$ = 0.89~mPa$\cdot$s. 
The fitted signals obtained using Eqs.~\ref{eq:signal_fitted} agree perfectly with the synthetic signal. For this special case of unit magnitude and zero phase transfer functions, the spanned signals obtained using Eqs.~\ref{eq:signal_spanned} also agree with the synthetic signal. The NRMSE values with respect to the ground-truth synthetic signal are 0.01\% and 0.1\% for the fitted signals at SNR = 10 and SNR = 1, respectively. It should be noted that the ground-truth synthetic signals differ at SNR = 10 vs. SNR = 1 due to differences in number of harmonics selected during SBR thresholding.

\begin{figure}[!t]
   \begin{center}
   \includegraphics{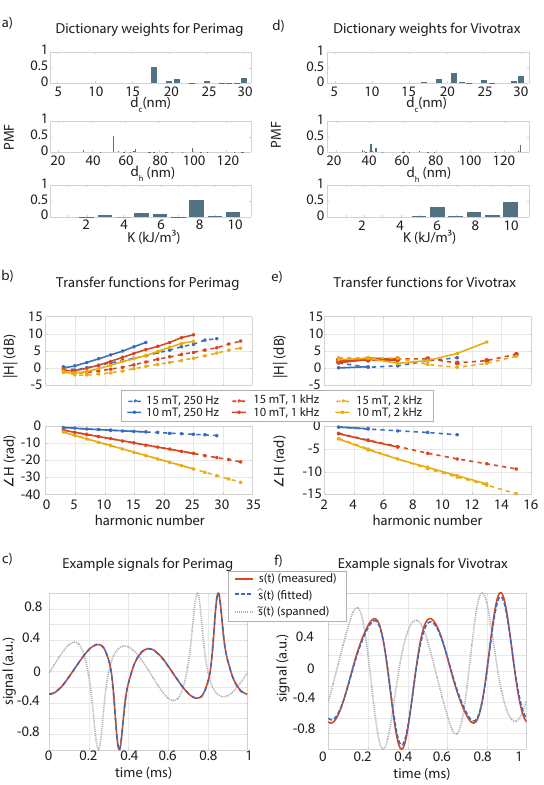} 
   \captionv{12}{}{Estimation and signal fitting results for Perimag and Vivotrax MNPs. Estimated weight vectors for (a) Perimag and (d) Vivotrax, displayed as PMFs for $d_c$, $d_h$, and $K$. Estimated magnitude and phase responses for transfer functions for the cases of (b) Perimag and (e) Vivotrax. Example measured, fitted, and spanned signals for (c) Perimag and (f) Vivotrax for $\eta$ = 0.89 mPa$\cdot$s at 10~mT, 1~kHz DF setting. The estimated transfer functions were normalized together, such that the 3$^{\textrm{rd}}$ harmonic for 10~mT, 250~Hz DF setting has 0~dB magnitude response. \label{fig:particle_properties}
    }  
    \end{center}
\end{figure}

\subsection{Estimation and Signal Fitting Results for MPS Measurements}
Figure~\ref{fig:particle_properties} shows the weight vector, transfer functions, and fitted signals of Perimag and Vivotrax, estimated using MPS measurements at 6 DF settings and 6 viscosity levels. The estimated weight vectors are depicted as PMFs for $d_c$, $d_h$, and $K$ in Figs.~\ref{fig:particle_properties}(a) and~\ref{fig:particle_properties}(d), for Perimag and Vivotrax, respectively. Given that both Perimag and Vivotrax are multicore nanoparticles, these distributions can be interpreted as the effective magnetic parameters that explain the received signals. 
The median values for the estimated parameters were $d_c$ = 18~nm, $d_h$ = 53~nm, and $K$ = 8~$kJ/m^3$ for Perimag, and $d_c$ = 21~nm, $d_h$ = 44~nm, and $K$ = 9~$kJ/m^3$ for Vivotrax. 

Figures~\ref{fig:particle_properties}(b) and~\ref{fig:particle_properties}(e) show the transfer functions representing the non-model based dynamics for Perimag and Vivotrax, respectively. Because Vivotrax samples had relatively lower signal than Perimag samples, fewer number of harmonics were included in its dictionary matrices and the transfer functions  (note the different x-axes ranges for Figs.~\ref{fig:particle_properties}(b) and~\ref{fig:particle_properties}(e)). While the amplitude responses for Vivotrax remain mostly flat, those for Perimag show a high-pass characteristics. This difference may stem from stronger non-model based dynamics for Perimag, such as dipolar interactions or non-uniaxial magnetic anisotropy. The phase responses display similar linear trends for both Perimag and Vivotrax, nearly independent of the DF amplitude. Given that linear phase indicates a time-domain delay, these results imply that the main cause of the phase response is the system delay from the measurement setup.

Figures~\ref{fig:particle_properties}(c) and~\ref{fig:particle_properties}(f) show example signals at 10 mT and 1 kHz DF setting, for $\eta$~=~0.89~mPa$\cdot$s. As expected from the phase responses, a similar time delay can be observed between the spanned and measured signals for both Perimag and Vivotrax. 
In addition, for Perimag, the width of the signal peak is slightly narrower for the fitted signal when compared to the spanned signal, due to the high-pass characteristics of the magnitude responses.

\subsection{Signal Prediction Results for MPS Measurements}
\begin{figure}[!t]
   \begin{center}
   \includegraphics{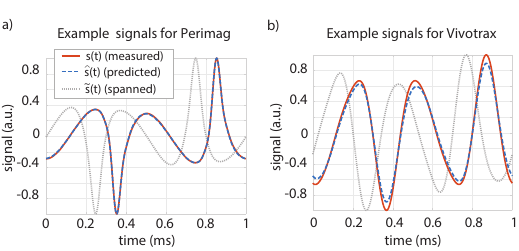} 
   \captionv{12}{}{Example signal prediction results for (a) Perimag and (b) Vivotrax, showing measured, predicted, and spanned signals for $\eta$ = 0.89~mPa$\cdot$s at 10~mT, 1~kHz DF settings. \label{fig:signal_prediction}
    }  
    \end{center}
\end{figure}

To validate the signal prediction performance of the proposed algorithm, the MPS measurements from each viscosity level was left out one by one, and the signal at that viscosity level was predicted as described in Eqs.~\ref{eq:signal_spanned_predicted} and~\ref{eq:signal_predicted}. Example signal prediction results are shown in Fig.~\ref{fig:signal_prediction} for Perimag and Vivotrax when the viscosity level of 0.89~mPa$\cdot$s is left out, for 10~mT and 1~kHz DF settings. The ground-truth measured signals are also displayed for reference. The spanned signals once again show a time shift with respect to the measured signal, whereas the predicted signals show near perfect temporal alignment, thanks to the estimated transfer function. Overall, the predicted Perimag signal shows a perfect agreement with the measured signal, while the predicted Vivotrax signal underestimates the signal peaks. Correspondingly, the NRMSE(\%) values are 0.26\% and 2.9\% for predicted Perimag and Vivotrax signals, respectively. 

\subsection{Quantitative Assessments for Signal Fitting and Signal Prediction}
\begin{figure}[!t]
   \begin{center}
   \includegraphics{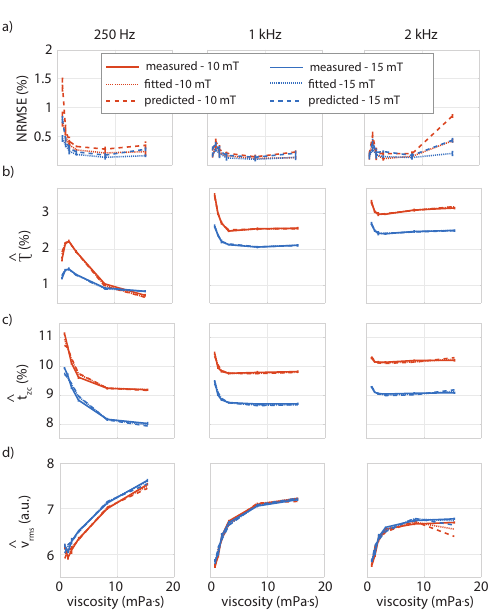} 
   \captionv{12}{}{Signal prediction and fitting performances for Perimag. Results at 3 different DF frequencies and 2 different DF amplitudes are plotted as functions of viscosity level. (a) NRMSE values with respect to the measured signal. (b) $\hat{\tau}$, (c) $\hat{t}_{zc}$, and (d) $\hat{v}_{rms}$ for measured, fitted, and predicted signals. The fitted signals utilized measurements at all DF settings and viscosity levels, whereas the predicted signal excluded all measurements at a given viscosity level when estimating the weight vector and transfer functions. \label{fig:perimag_metrics}
    }  
    \end{center}
\end{figure}

\begin{figure}[!t]
   \begin{center}
   \includegraphics{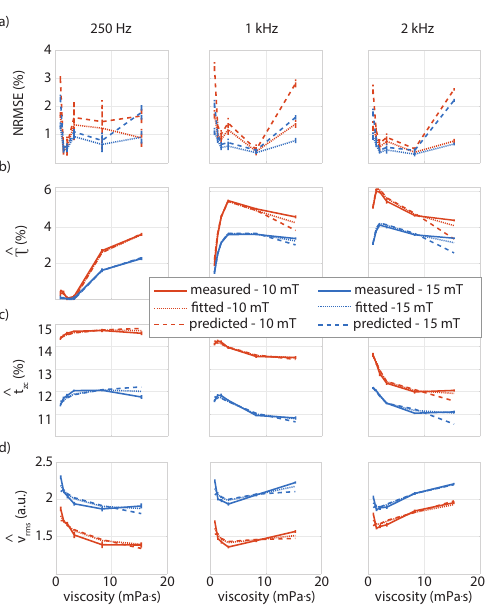} 
   \captionv{12}{}{Signal prediction and fitting performances for Vivotrax. Results at 3 different DF frequencies and 2 different DF amplitudes are plotted as functions of viscosity level. (a) NRMSE values with respect to the measured signal. (b) $\hat{\tau}$, (c) $\hat{t}_{zc}$, and (d) $\hat{v}_{rms}$ for measured, fitted, and predicted signals. 
    The fitted signals utilized measurements at all DF settings and viscosity levels, whereas the predicted signal excluded all measurements at a given viscosity level when estimating the weight vector and transfer functions.  \label{fig:vivotrax_metrics}
    }  
    \end{center}
\end{figure}

The results of the quantitative evaluations for fitted and predicted signals with respect to the measured signals are presented in Fig.~\ref{fig:perimag_metrics} and~\ref{fig:vivotrax_metrics} for Perimag and Vivotrax, respectively. The metrics are plotted as functions of the viscosity level, showing the performances at 3 different DF frequencies and 2 different DF amplitudes. The fitted signals at a given viscosity were calculated as described in Eq.~\ref{eq:signal_fitted}, using the measurements at all DF settings and viscosity levels. The predicted signals were calculated as described in Eq.~\ref{eq:signal_predicted}, after excluding all measurements at a given viscosity level.

\textbf{1) Results for Perimag:} For Perimag, Fig.~\ref{fig:perimag_metrics}(a) shows that the NRMSE(\%) values remain below 0.92\% and 1.51\% for fitted and predicted signals, respectively. The highest errors occur at the extreme experimental conditions, i.e., at the lowest $\eta$ and $f_d$, and the highest $\eta$ and $f_d$. At the intermediate DF frequency of 1~kHz, the NRMSE remains below 0.43\%. 
Overall, these low NRMSE values indicate a high fidelity match with the measured signal.

Figures~\ref{fig:perimag_metrics}(b)-(d) compare $\hat{\tau}$, $\hat{t}_{zc}$, and $\hat{\text{v}}_{rms}$ values for measured, fitted, and predicted signals. Overall, these metrics display a strong dependence on viscosity at 250~Hz, whereas the trends flatten as the DF frequency is increased to 2~kHz. These trends are consistent with previous work in the literature that has shown reduced sensitivity to viscosity at higher DF frequencies~\cite{Utkur2017,Utkur2022}. Importantly, these trends are successfully captured by both the fitted and predicted signals. Comparing the metrics from measured, fitted, and predicted signals, one can see an excellent match at all viscosity levels, particularly for $\hat{\tau}$ and $\hat{t}_{zc}$. 
 
\textbf{2) Results for Vivotrax:} Figure~\ref{fig:vivotrax_metrics}(a) illustrates the NRMSE(\%) values of fitted and predicted signals for Vivotrax, which remain below 1.8\% and 3.5\%, respectively. The highest NRMSE values are observed at extreme viscosities at all DF frequencies. Notably, the NMRSE values for Vivotrax are higher than those of Perimag, which can be attributed to the lower SNR of Vivotrax measurements. 

Figures~\ref{fig:vivotrax_metrics}(b)-(d) compare $\hat{\tau}$, $\hat{t}_{zc}$, and $\hat{\text{v}}_{rms}$ for the measured, fitted, and predicted signals. In general, $\hat{\tau}$ and $\hat{t}_{zc}$ values are larger than those of Perimag, indicating more substantial relaxation effects for Vivotrax. 
In addition, $\hat{\text{v}}_{rms}$ is approximately 4-fold smaller than that of Perimag, indicating significantly reduced signal efficiency for Vivotrax. Furthermore, unlike in Perimag, these metrics for Vivotrax maintain a strong dependence on viscosity even at higher DF frequencies.
Interestingly, the trends for Vivotrax at 2~kHz are similar to the trends exhibited by Perimag at 250~Hz. This observation implies that Vivotrax experiences similar magnetization dynamics with Perimag but at higher DF frequencies, a result consistent with the literature~\cite{Utkur2017}. Importantly, these trends are successfully captured by both the fitted and predicted signals, albeit with slightly larger errors for the case of Vivotrax than for Perimag.

\begin{figure}[!t]
   \begin{center}
   \includegraphics{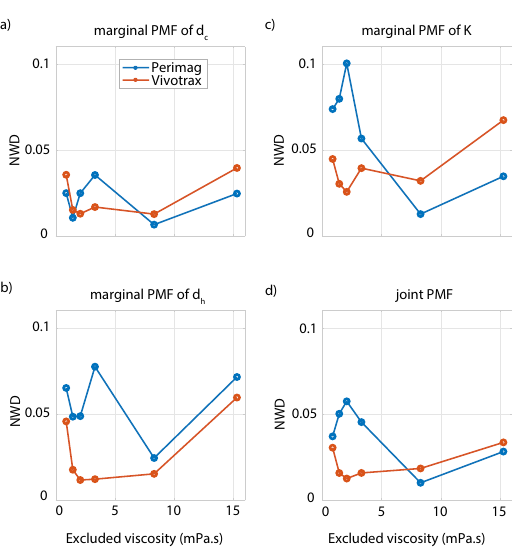} 
   \captionv{12}{}{Normalized Wasserstein distances (NWDs) for Perimag and Vivotrax, assessing how the estimated weight vectors are affected when all measurements at a given viscosity level are excluded. NWD values for marginal PMFs of (a) $d_c$, (b) $d_h$, and (c)$K$, and for (d) joint PMF. \label{fig:NW}
    }  
    \end{center}
\end{figure}

\subsection{Quantitative Assessments for Dictionary Weights}
The NWD metric was utilized to assess how the estimated dictionary weights are affected by the process of excluding all measurements from a given viscosity level. The weight vectors obtained by including all available measurements were utilized as references (i.e., the PMFs in Figs.~\ref{fig:particle_properties}(a) and ~\ref{fig:particle_properties}(d) for Perimag and Vivotrax, respectively).  

NWDs for marginal PMF of $d_c$, $d_h$, and $K$ as well as their joint PMF are shown in Fig.~\ref{fig:NW}, as functions of the excluded viscosity level. Overall, NWD remains below 0.10 and 0.07 for Perimag and Vivotrax, respectively. Specifically, for the marginal PMF of $d_c$, NWD remains below 0.03 and 0.04 for Perimag and Vivotrax, respectively. The viscosity levels that, when excluded, lead to a relatively larger deviation in the estimated weight vector differ for Perimag and Vivotrax. For Vivotrax, including the highest and lowest viscosity levels is more important to obtain weights that are closer to the reference. In contrast, including the low-to-mid viscosity levels are more important for Perimag. To illustrate, for the joint PMFs, the highest NWD of 0.08 occurs at 2.08~mPa$\cdot$s for Perimag, and the highest NWD of 0.05 occurs at 15.33~mPa$\cdot$s for Vivotrax. Nonetheless, the NWD values remain low for all cases, indicating robustness in dictionary weight estimation.

\section{Discussion}
In this work, we have proposed a novel calibration-free dictionary-based algorithm for MNP signal prediction. In addition to the model-based dynamics, the proposed approach incorporates the non-model based dynamics by modeling them as an LTI system, and jointly estimates the dictionary weight vector and transfer functions. The predicted and measured signals at all viscosity levels demonstrate an excellent match, successfully capturing the trends as a function of viscosity.

\subsection{Magnetization Model}
This work utilized a coupled Brown-Néel rotation model to construct the dictionaries. As shown in Fig.~\ref{fig:model_comparison}, the coupled model captures unique magnetization characteristics compared to simpler models.
The MPS experiments further reaffirm the necessity of utilizing the coupled model, as they demonstrate that the MNPs experience both Brown and Néel rotations for the range of DF settings utilized in this work. As illustrated in Fig.~\ref{fig:perimag_metrics} for Perimag, lower DF frequencies display a strong viscosity sensitivity that implies Brown rotation taking place. In contrast, the viscosity sensitivity diminishes as DF frequency increases to 2~kHz, suggesting that the Brown rotation becomes less effective and is superseded by Néel rotation. 

Because the pure Néel model and the Langevin model do not incorporate the viscosity effect, they are not suitable for the purposes of this study. To test whether the pure Brown model would suffice in capturing the trends, we formed dictionaries by using this model for the same set of magnetic parameters. Since the pure Brown model does not incorporate the anisotropy effect, $K$ was not included as a parameter~\cite{Shasha2020}. The signal prediction performance using the pure Brown model was poor, yielding fitted and predicted signals that both diverged from the measured signals (results not shown).

\subsection{Non-Model-Based Dynamics}
While it provides a powerful framework for understanding MNP dynamics, the employed coupled Brown-Néel rotation model is not without limitations. It assumes uniaxially symmetric particles with uniaxial magnetic anisotropy~\cite{Weizenecker2018}. Anisotropy stemming from the cubic crystal symmetry is ignored due to the relatively low energy barrier in the crystal~\cite{Weizenecker2012}. 
Furthermore, the coupled model assumes single-core particles and does not account for the interactions among multi-core particles. Here, multi-core Perimag and Vivotrax MNPs were treated as single-core particles with an effective $d_c$, as commonly done in the literature~\cite{Eberbeck2013,Yoshida2013,Ota2019}.  

The proposed algorithm hypothesizes that the effects of non-model-based dynamics can be formulated as an LTI system that accounts for (1) the magnetization dynamics that are not accounted for by the magnetization model and (2) the transfer function of the measurement setup. The LTI assumption holds for the measurement setup, since it is composed of linear circuit components. 

The LTI assumption for the non-model-based magnetization dynamics, such as magnetodipolar interactions and dynamics resulting from nonuniaxial magnetic anisotropy, deserves a closer look. For practical purposes such as enabling signal prediction at different viscosities, the formulation in Eq.~\ref{eq:overall_problem} assumes that these dynamics are dependent on the DF settings, but not on the viscosity level.
The quantitative and qualitative evaluations conducted on fitted and predicted signals in Figs.~\ref{fig:signal_prediction}-\ref{fig:vivotrax_metrics} confirm the effectiveness of this approach for the employed range of DF settings and viscosities. Nonetheless, the relatively reduced fitting and prediction performances at the highest/lowest viscosities in Figs.~\ref{fig:perimag_metrics}-\ref{fig:vivotrax_metrics} may imply that this assumption may not be fully valid for a wider range of viscosities. 

The results in Fig.~\ref{fig:synt_results} validate the success of the proposed algorithm utilizing synthetic signals. When synthetic signals were generated using the same parameters in Table~\ref{tab:MagProperties_united} but for transfer functions with polynomial amplitudes and linear phases, the algorithm was still successful in jointly estimating the weight vector and transfer functions (results not shown).

In Fig.~\ref{fig:particle_properties}, the estimated transfer functions for Perimag show high-pass characteristics, implying that non-model-based magnetization dynamics contribute to improving the MPI resolution of Perimag. This observation is consistent with the literature reporting improved MPI resolution for this MNP due to increased dipolar interactions~\cite{Eberbeck2011,Tay2021_SF}. Note that if the employed magnetization model could fully capture the magnetization dynamics, the estimated transfer functions would solely represent those of the measurement setup, and would therefore be identical for different MNPs. While the system transfer function could be determined via additional calibration measurements~\cite{Thieben2023}, the proposed algorithm offers the advantage of not requiring any calibrations. 

\subsection{Magnetic Parameters}
The proposed algorithm relies on a dictionary matrix of simulated MNP signals. It should be noted that particles with different magnetic parameters may display similar magnetization responses, which in turn can make the inverse problem ill-conditioned. This problem can be further exacerbated if the measured signal has low SNR. In such a case, fewer harmonics survive SBR thresholding, causing the system of equations to be highly underdetermined. One potential approach to solve this problem could be to optimize the range and sampling of the magnetic parameters based on the SNR level, with the goal of improving the conditioning of the dictionary matrix. A detailed analysis on how the SNR level and the selection of the magnetic parameters affect the accuracy of signal prediction remains an important future work.

\subsection{Computation Time}
Simulating the dictionary matrix that is at the heart of the proposed method requires extensive computation time. For example, solving the ODE underlying the coupled Brown-Néel model can take anywhere from minutes to a few days for a single particle, depending on its magnetic parameters. Particularly, it takes longer to simulate the magnetic responses for particles with large $d_c$ and large $K$. Here, we have utilized parallelization on a CPU to simultaneously simulate particles with different parameters. Further reduction in computation time could be achieved using GPUs and/or faster ODE solvers \cite{Karimov2017}. Alternatively, deep learning based approaches can be utilized for simulating the magnetization response \cite{Knopp2023}.  

\subsection{Other Applications of Signal Prediction}
One important application of the proposed algorithm is to predict the signal trends in a wide range of viscosities from MPS measurements performed at only a few different viscosity levels. A similar approach could also be followed for predicting the trends as a function of temperature. The proposed framework can also be extended to MPI signal prediction for applications such as viscosity mapping and temperature mapping, to help determine the optimal DF settings or optimal MNP type. Note that such an extension would require simulations that incorporate not only the DF but also the selection field of the MPI system.

\section{Conclusion}
In this work, we proposed a calibration-free iterative dictionary-based algorithm that jointly estimates the dictionary weights and non-model based dynamics, allowing MNP signal predictions at untested settings. The quantitative and qualitative assessments on synthesized and measured signals demonstrate the accuracy of this approach in predicting the trends in MNP signals as a function of viscosity. The proposed approach offers numerous potential applications, such as determining the optimal DF settings or optimal MNPs for viscosity mapping or temperature mapping using MPI.

\section*{Acknowledgments}
\addcontentsline{toc}{section}{\numberline{}Acknowledgments}
This work was supported by the Scientific and Technological Research Council of Turkey (TUBITAK) under Grant 120E208. The authors would like to thank Prof. Jürgen Weizenecker for the invaluable discussions on the coupled Brown-Néel model, and to Prof. Tolga Çukur for computational support. A preliminary version of this work was presented in IWMPI 2022. 

\section*{References}
\addcontentsline{toc}{section}{\numberline{}References}
\vspace*{-20mm}

\bibliography{references}
\bibliographystyle{./medphy.bst}

\end{document}